# Optimizing Traversing and Retrieval Speed of Large Breached Databases

Mayank Gite
CyberSapiens United LLP, India
mayank@cybersapiens.in

## Abstract

Breached data refers to the unauthorized access, theft, or exposure of confidential or sensitive information. Breaches typically occur when malicious actors or unauthorized users breach secure systems or networks, resulting in compromised personally identifiable information (PII), protected or personal health information (PHI), payment card industry (PCI) information, or other sensitive data. Data breaches are often the result of malicious activities such as hacking, phishing, insider threats, malware, or physical theft. The misuse of breached data can lead to identity theft, fraud, spamming, or blackmailing. Organizations that experience data breaches may face legal and financial consequences, reputational damage, and harm to their customers or users. Breached records are commonly sold on the dark web or made available on various public forums. To counteract these malicious activities, it is possible to collect breached databases and mitigate potential harm. These databases can be quite large, reaching sizes of up to 150 GB or more. Typically, breached data is stored in the CSV (Comma Separated Value) format due to its simplicity and lightweight nature, which reduces storage requirements. Analyzing and traversing large breached databases necessitates substantial computational power. However, this research explores techniques to optimize database traversal speed without the need to rent expensive cloud machines or virtual private servers (VPS). This optimization will enable individual security researchers to analyze and process large databases on their personal computer systems while significantly reducing costs.

**Keywords - Breach, Records, Databases, Analyze**

## 1 Introduction

The move towards the internet by people, businesses, and organizations alike has been a revolutionary phenomena in recent years. The internet has become more and more integrated into our daily lives as a result of advancements in technology and better connectivity. People now use the internet for a wide range of activities, such as communication, information access, entertainment, and online shopping. It has brought about unprecedented convenience, allowing people to communicate with people around the world, instantly access large amounts of information, and complete tasks effectively. This gives malicious actors an open playground and a huge opportunity. Recently, we read about how big organizations get hacked, resulting in a Data Breach. Breached data refers to any confidential or sensitive information that has been unauthorizedly stolen, accessed, or exposed. Breach data can consist of PII (personally identifiable information), PHI (protected or personal health information), PCI (payment card industry) data, or any other sensitive information. Later, this data is sold on Dark markets, or made publicly available on various forums. Breached data can be used for malicious purposes such as identity theft, fraud, spamming, or blackmailing. These activities can be mitigated in various ways if we process and analyze breached databases. Breached databases are sometimes as big as 150GB and can exceed even more, which causes challenges for an individual to process this data. These databases are generally stored in CSV (Comma Separated Value) format. CSV provides simplicity and is lightweight, thus reducing storage. One of the easiest ways to process big databases is to rent Cloud Computing machines, or VPS (virtual private servers), which is a costly approach. In this paper, I will discuss different optimized approaches through which any individual security researcher can process such large Breached Records with minimal computing power, making it cost- and resource-effective.

## 2 Approaches

Tested on system with configuration:

- Processor: Ryzen 7 4800H
- RAM: 8GB
- Storage Media: NVME SSD
- Operating System: GNU/Linux

Information of breached database used:

- Size: 21.3GB
- Format: CSV
- No of Values in a row: 59
- No of Columns: 56,329,605
- Total No of Records: 56,329,605.

### 2.1 GNU/grep

GNU/grep is a utility for Unix-like Operating Systems that searches one or more input files for lines containing a match to a specified pattern. By default, GNU grep outputs the matching lines. GNU grep uses various algorithms that make string or pattern searching, matching, and finding extremely fast. These algorithms include the Boyer-Moore algorithm, which is a fast string searching algorithm[1]; the Aho AV algorithm for finding patterns in strings and efficient string matching[2]; and others[3] Any input provided to grep is treated as a string. GNU grep typically searches for strings line by line within a file. It reads the input sequentially and processes it line by line. Disadvantages of grep are: slow search time, inability to perform various operations on data, and unsorted data. Following are calculated GNU grep benchmarks:

| Email ID | Approx Row No | Time taken (in seconds) |
|---|---|---|
| Email ID 1 | 1/4th i.e 13,573,500 | 18.5890 |
| Email ID 2 | 2/4th i.e 27,147,000 | 16.7470 |
| Email ID 3 | 3/4th i.e 40,720,500 | 16.7370 |
| Email ID 4 | 4/4th i.e 54,294,000 | 16.9150 |
| Invalid Email | Invalid Row | 16.8300 |
| **Average Time** | | **17.1636** |

Table 1.0: Time taken to find record using Email ID.

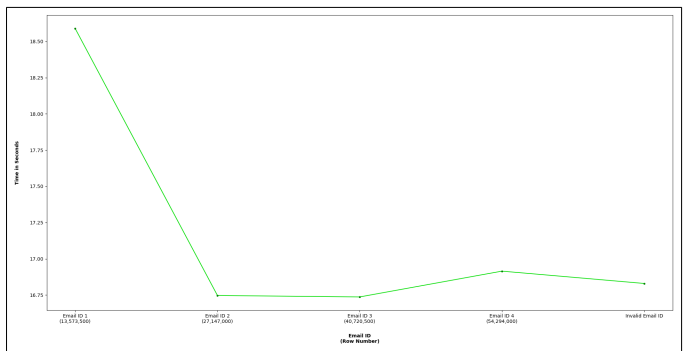

Graph 1.0: GNU/grep Benchmarks.

## 2.2 Python Pandas

pandas is a fast, powerful, flexible, and easy-to-use open-source data analysis and manipulation tool built on top of the Python programming language [4]. Pandas is built on top of two primary data structures: DataFrame and Series. A Series is a one-dimensional object that resembles an array and may store any type of data. It consists of a sequence of values and associated labels, referred to as the index. The index works similarly to a dictionary, allowing fast and efficient access to the data. It is implemented using data structures like hash tables or search trees, depending on the type of index used. A DataFrame is a two-dimensional table-like data structure consisting of rows and columns. Conceptually, it resembles a spreadsheet or a SQL table. A data frame can hold different types of data and provide labelled axes (rows and columns) for easy indexing and manipulation. In order to manipulate, analyze, and transform data, Pandas makes use of these data structures and offers a variety of functions and methods. It takes advantage of vectorized operations, which perform computations on entire arrays rather than individual elements, resulting in faster execution times. Data stored in a DataFrame or Series is directly stored in volatile memory for processing and manipulation. To mitigate this problem, pandas provides the chunksize parameter in the read_csv() function, which loads large datasets in smaller and more manageable chunks. Depending on the particular operations carried out and the size of the data chunks being processed, it is still possible to run into memory restrictions due to Cumulative Memory Usage.

Due to High Memory consumption, I extracted small chunk from large breached dataset and performed benchmarks on it:

- Size: 262 MB
- Format: CSV
- No of Values in a row: 59
- No of Rows: 674,918
- Total No of Records: 39,820,162.

Memory Usage for processing 262MB dataset chunk was 285.8MB. Therefore memory required to process primary dataset (21.5GB) can be calculated as following:

$$Ratio = Memory\ Usage\ /\ File\ Size = 285.8\ MB\ /\ 262\ MB$$
$$Memory\ Usage = Ratio * File\ Size = Ratio * 21.5\ GB$$

Converting the file size to megabytes (MB):
$$21.5\ GB = 21.5 * 1024\ MB = 22,118.4\ MB$$

Substituting the values into the equation:
$$Memory\ Usage = (285.8\ MB\ /\ 262\ MB) * 22,118.4\ MB$$

Simplifying the expression:
$$Memory\ Usage = (285.8 * 22,118.4)\ /\ 262\ MB$$

Calculating the result:
$$RAM\ USAGE \approx 24,146.74\ MB$$

Time taken to read 262MB dataset chunk was 4.740 seconds. Therefore time required to read primary dataset (21.5GB) can be calculated as following:

Convering file sizes to same unit(GB):
$$262\ MB = 0.262\ GB$$
$$21.5\ GB = 21.5\ GB$$

Setting up proportion:
$$Processing\ time\ (seconds)\ /\ File\ size\ (GB) = Processing\ time\ (seconds)\ /\ File\ size\ (GB)$$
$$4.740\ secs\ /\ 0.262\ GB = x\ secs\ /\ 21.5\ GB$$

Finding estimated processing time for the 21.5 GB file:
$$(4.740\ secs * 21.5\ GB)\ /\ 0.262\ GB \approx 389.966\ secs$$

Note: Memory usage and processing time might increase as size of dataset increases.

Following are the calculated benchmarks using pandas:

| Appox. row possition in Table | Time taken (in seconds) |
|---|---|
| 1/4th i.e 13,573,500 | 0.0010 |
| 2/4th i.e 27,147,000 | 0.0010 |
| 3/4th i.e 40,720,500 | 0.0010 |
| 4/4th i.e 54,294,000 | 0.0010 |
| **Average Time** | **0.0010** |

Table 2.0: Time taken to find record using row number

| Email ID | Time taken (in seconds) |
|---|---|
| Email ID 1 | 0.1713 |
| Email ID 2 | 0.1729 |
| Email ID 3 | 0.1716 |
| Email ID 4 | 0.1705 |
| **Average Time** | **0.1715** |

Table 2.1: Time taken to find record using Email ID.

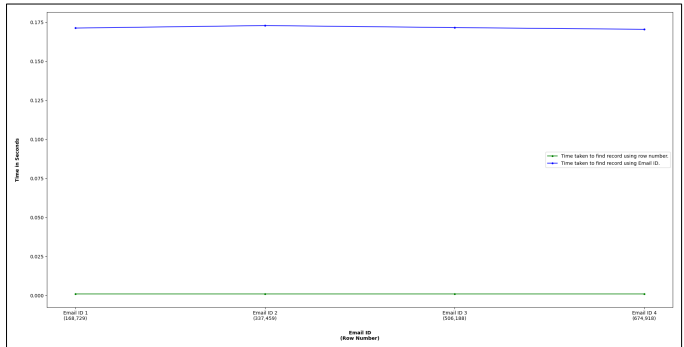

Graph 2.0: Python pandas Benchmarks.

## 2.3 RDBMS (MySQL):

MySQL is an open-source relational database management system (RDBMS) that is widely used for managing and organizing structured data [5]. MySQL allows users to store, retrieve, and manipulate data using the SQL (Structured Query Language) programming language. It provides a robust and scalable platform for creating and managing databases, handling data transactions, and implementing data security.

### Importing CSV into MySQL:

MySQL supports importing data from CSV files. However, before performing the import, it is crucial to create a corresponding MySQL table that mirrors the structure of the data within the CSV. It is important to carefully choose DataType and its length for columns, as they determine the size and efficiency of the table. Adding PRIMARY or UNIQUE constraints increases query performance drastically, though it is difficult to apply these constraints to tables as the majority of breached databases contain duplicate records (unless databases are cleaned). MySQL provides the "LOAD DATA" statement, which reads rows from a text file into a table at a very high speed. It efficiently processes and inserts data in bulk, significantly reducing the import time compared to iterative insertion methods. By adding Field and Line Handling clauses in the "LOAD DATA" statement, importing options can be configured. "LOAD DATA" statement, simplifies and accelerates the data import process.

### Table information after importing CSV:

- Size of Table after import: 24.1GB
- Type: InnoDB
- Overhead Size: 5.0MB
- Number of Rows: 56,329,605
- Total No of Records: 3,202,746,000

## 2.3.1 MySQL (Default Config):

InnoDB is the default Storage Engine used by MySQL. It is intended to offer high-performance transactional capabilities and ACID compliance. It supports concurrent access with row-level locking, ensures data integrity through ACID compliance, and employs multi-versioning concurrency control (MVCC) for improved scalability. With features like crash recovery, foreign key constraints, and a buffer pool for effective data caching, InnoDB offers durability, reliability, and optimal performance. It is the default and widely recommended MySQL storage engine due to its ability to manage complex applications and preserve consistent and dependable data. Following are calculated benchmarks using MySQL's default configuration:

| Appox. row possition in Table | Time taken (in seconds) |
|---|---|
| 1/4th i.e 13,573,500 | 44.8576 |
| 2/4th i.e 27,147,000 | 89.5477 |
| 3/4th i.e 40,720,500 | 134.3216 |
| 4/4th i.e 54,294,000 | 179.1416 |
| Invalid Row | 183.6712 |
| **Average Time** | **126.3079** |

Table 3.1.0: Time taken to find record using row number.

| Email ID | Time taken (in seconds) |
|---|---|
| Email ID 1 | 185.4700 |
| Email ID 2 | 119.0708 |
| Email ID 3 | 186.7824 |
| Email ID 4 | 121.5162 |
| Invalid Email | 186.2700 |
| **Average Time** | **159.8215** |

Table 3.1.1: Time taken to find record using Email ID.

| Appox. row possition in Table | Time taken (in seconds) |
|---|---|
| 1/4th i.e 13,573,500 | 40.3465 |
| 2/4th i.e 27,147,000 | 81.3524 |
| 3/4th i.e 40,720,500 | 122.2772 |
| 4/4th i.e 54,294,000 | 163.0196 |
| Invalid Row | 169.7557 |
| **Average Time** | **115.3503** |

Table 3.1.2: Time taken to find record with DataType Integer, using row number.

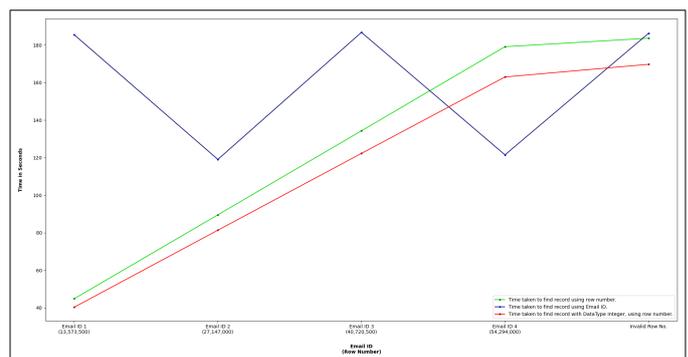

Graph 3.0: MySQL(Default Config) Benchmarks.

## 2.3.2 MySQL (Changing storage engine to MyISAM):

MyISAM is a storage engine in MySQL. It is based on the older ISAM storage engine (no longer available) but has many useful extensions. MyISAM storage engine creates two separate files for each table:

- MyISAM Data File (.MYD), which stores table definitions and data records in a structured format that allows efficient read and write operations,
- MyISAM Index File (.MYI) contains index data for the table, which allows efficient data retrieval based on indexed columns.

MyISAM offers several advantages in terms of read speed and performance. Its efficiency stems from features such as Table-level Locking, which allows multiple read operations to occur simultaneously. This concurrency enhances the overall performance of the database; Simpler Data Structures, which enable faster data access and retrieval; No Transaction Overhead, This makes it particularly suitable for situations where read operations significantly outnumber write operations and Full-Text Search Optimization making it a preferred choice for applications requiring efficient text search capabilities. Due to its focus on read operations, MyISAM is particularly suitable for handling breached records. It excels in scenarios where data retrieval takes precedence, ensuring efficient and swift access to information. By leveraging these capabilities, MyISAM becomes a reliable choice for applications that prioritize fast read speed and performance optimization.

**Table information after importing CSV:**

- Size of Table after import: 20.2GB
- Type: MyISAM
- Index Size: 1.0 KB
- Overhead Size: 0B
- Number of Rows: 56,329,605
- Total No of Records: 3,202,746,000

Following are the calculated benchmarks using MySQL with MyISAM storage engine:

| Appox. row possition in Table | Time taken (in seconds) |
|---|---|
| 1/4th i.e 13,573,500 | 11.6490 |
| 2/4th i.e 27,147,000 | 18.7989 |
| 3/4th i.e 40,720,500 | 26.4822 |
| 4/4th i.e 54,294,000 | 34.3418 |
| Invalid Row | 35.4401 |
| **Average Time** | **25.3434** |

Table 3.2.0: Time taken to find record using row number.

| Email ID | Time taken (in seconds) |
|---|---|
| Email ID 1 | 33.0233 |
| Email ID 2 | 33.8236 |
| Email ID 3 | 33.8813 |
| Email ID 4 | 33.6320 |
| Invalid Email | 33.4518 |
| **Average Time** | **33.5624** |

Table 3.2.1: Time taken to find record using Email ID.

| Appox. row possition in Table | Time taken (in seconds) |
|---|---|
| 1/4th i.e 13,573,500 | 7.8218 |
| 2/4th i.e 27,147,000 | 15.7970 |
| 3/4th i.e 40,720,500 | 23.7650 |
| 4/4th i.e 54,294,000 | 31.1407 |
| Invalid Row | 32.0559 |
| **Average Time** | **22.1160** |

Table 3.2.2: Time taken to find record with DataType Integer, using row number.

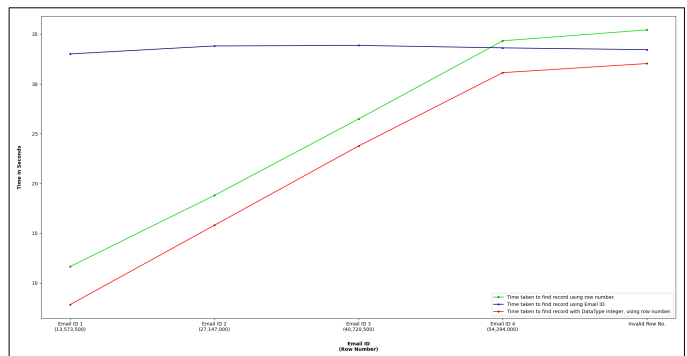

Graph 3.1: MySQL(Storage Engine MyISAM) Benchmarks

### 2.3.3 MySQL with storage engine MyISAM (Adding Index):

An index acts as a data structure that enhances the speed of data retrieval operations on database tables. It is created on one or more columns of a table to facilitate fast searching and sorting of data. When an index is created on a column or set of columns, MySQL constructs an internal data structure, typically a Binary-tree or hash. These structures establish a link between the indexed column values and the physical location of the associated rows in the table. This connection allows the database engine to locate and retrieve the required data more efficiently. Such optimization is crucial for managing large datasets and ensuring prompt query execution.

**Table information after importing CSV:**

- Size of Table after import: 20.2GB
- Type: MyISAM
- Index Size: 936.7MB
- Overhead Size: 0B
- Total Size: 21.1GB
- Number of Rows: 56,329,605
- Total No of Records: 3,202,746,000

Following are calculated benchmarks using MySQL with the MyISAM storage engine and indexes (created on columns "Contact Number" and "Email Address"):

| Appox. row possition in Table | Time taken (in seconds) |
|---|---|
| 1/4th i.e 13,573,500 | 8.0653 |
| 2/4th i.e 27,147,000 | 14.9719 |
| 3/4th i.e 40,720,500 | 22.7882 |
| 4/4th i.e 54,294,000 | 30.7508 |
| Invalid Row | 31.9110 |
| **Average Time** | **21.6974** |

Table 3.3.0: Time taken to find record using row number.

| Email ID | Time taken (in seconds) |
|---|---|
| Email ID 1 | 0.0005 |
| Email ID 2 | 0.0100 |
| Email ID 3 | 0.0060 |
| Email ID 4 | 0.0086 |
| Invalid Email | 0.0004 |
| **Average Time** | **0.0047** |

Table 3.3.1: Time taken to find record using Email ID.

| Appox. row possition in Table | Time taken (in seconds) |
|---|---|
| 1/4th i.e 13,573,500 | 1.7371 |
| 2/4th i.e 27,147,000 | 3.6439 |
| 3/4th i.e 40,720,500 | 5.3739 |
| 4/4th i.e 54,294,000 | 7.1228 |
| Invalid Row | 7.4099 |
| **Average Time** | **5.0575** |

Table 3.3.2: Time taken to find record with DataType Integer, using row number.

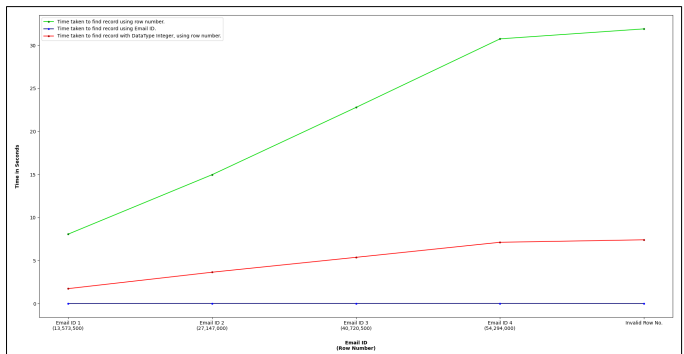

Graph 3.2: MySQL with storage engine MyISAM(Adding Index)

## 3 Overall Benchmark

| Approach | Average Time (in seconds) |
|---|---|
| GNU/grep | 17.1600 |
| Python pandas | 0.1715 |
| MySQL(default config) | 159.8215 |
| MySQL(storage engine MyISAM) | 33.5624 |
| MySQL(storage engine MyISAM with Index) | 0.0047 |

Table 4.0: Overall benchmark for finding records using Email ID.

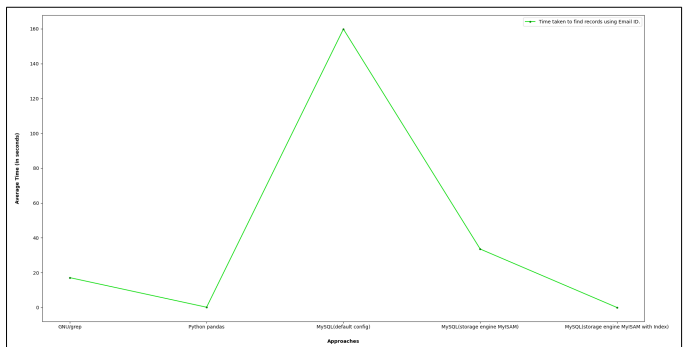

Graph 4.0: Overall benchmark for finding records using Email ID.

## Conclusion

In conclusion, this research paper has examined the important issue of processing large breached databases on personal computers or local systems efficiently for security researchers. Through a comprehensive analysis, I investigated various techniques and strategies to enhance the efficiency and speed of data traversal and retrieval in computer systems, specifically focusing on the context of handling large breached databases. The findings of this research underscore the significance of optimizing data processing for security researchers dealing with large breached databases on their personal computers. By leveraging advanced algorithms, parallel processing techniques, and efficient data structures, it is possible to achieve efficient and timely data traversal and retrieval, even on personal computer setups. The contributions of this study extend beyond theoretical advancements, as they directly address the practical challenges faced by security researchers. By enabling efficient processing of large breached databases on personal computers, researchers can reduce their reliance on expensive infrastructure, such as VPS or cloud computing systems. This not only leads to cost savings but also provides researchers with greater control and flexibility over their data processing tasks. The outcomes of this study contribute to empowering security researchers with enhanced capabilities and cost-effective solutions, ultimately aiding in the detection and mitigation of security breaches. Moreover, this research opens avenues for further exploration and refinement of techniques to handle even larger datasets and optimize resource utilization.